\documentclass[11pt,twoside]{article}

\usepackage{asp2006}
\usepackage{epsf}
\usepackage{psfig}
\usepackage{lscape}

\begin{document}
\title{SED modeling of Young Massive Stars}   %%% Fill in title

\author{Thomas P. Robitaille}   %%% Fill in author names
\affil{SUPA, School of Physics and Astronomy, University of St Andrews, North 
Haugh, KY16 9SS, St Andrews, United Kingdom}    %%% Fill in author affiliations

\begin{abstract} %%% Abstract to run on from here.
In this contribution, I review the applications and potential limitations of the spectral energy distribution fitting tool that I have developed, with a strong emphasis on the limits to which this tool can be used to improve our understanding of massive star formation. I discuss why our current grid of models cannot be used to distinguish between the several competing theories of massive star formation. I also discuss stellar mass determinations, artificial correlations between parameters in the grid of models, multiplicity, confusion, dust assumptions, and unique fits. I briefly review the improvements we intend to carry out for our next grid of models, which will eliminate many of these limitations. Finally, I show examples of applications of this tool to massive young stars.

\end{abstract}

%%% MAIN BODY OF TEXT GOES HERE. CONSULT "INSTRUCTIONS FOR AUTHORS USING
%%% LATEX2E MARKUP", SECTIONS 2.3-2.6 FOR HELP WITH EQUATIONS, FIGURES,
%%% AND TABLES.

%\section{}   %%% Top level section head (remove "%" symbol)
%\subsection{}   %%% Second level section head (remove "%" symbol)
%\subsubsection{}   %%% Lowest level section head (remove "%" symbol)
%\section*{}    %%% Unnumbered top level section head (remove "%" symbol)
%\subsection*{}   %%% Unnumbered second level section head (remove "%" symbol)

%%%%%%%%%%%%%%%%%%%%%%%%%%%%%%%%%%%%%%%%%%%%%%%%%%%%%%%%%%%%%%%%%%%%%%%%%%%%%%%%%%%%%%%%%%%%%%
\section{Introduction}
%%%%%%%%%%%%%%%%%%%%%%%%%%%%%%%%%%%%%%%%%%%%%%%%%%%%%%%%%%%%%%%%%%%%%%%%%%%%%%%%%%%%%%%%%%%%%%

In Robitaille et al. (2006, hereafter R06), we presented a large grid of model spectral energy distributions (SEDs) for young stellar objects (YSOs) for a range of evolutionary stages (from embedded protostars to optically thin disks) and stellar masses (from 0.1M$_\odot$ to 50M$_\odot$), and we used this grid of models to explore the relation between the physical conditions in young stars and their colours at near- and mid-infrared wavelengths. We then presented a linear regression SED fitting technique (Robitaille et al., 2007, hereafter R07) which we used to analyse the SEDs of young stars in the Taurus-Auriga star-formation region, as a proof of concept that this technique can correctly identify the physical conditions in YSOs. The tool has subsequently been used for a number of studies, including an analysis of the YSO population in M16 (Indebetouw et al., 2007), the IRAS 18507+0121 star formation region (Shepherd et al., 2007), the Rosette Molecular Cloud (Poulton et al., 2007), and star formation regions in the SMC (Simon et al., 2007) and LMC (Whitney et al., 2008). The models and the fitting tool are available to the community through a web interface\footnote{http://www.astro.wisc.edu/protostars}. The web-based tool has already been used by various groups to model massive YSOs, and some of their results were presented at this meeting.\\

Since this tool is being widely used to analyse the SEDs of low, intermediate, and high-mass YSOs, I use this contribution to re-iterate the main caveats and limitations that apply when using this tool (\S\ref{limit}), to discuss planned improvements in the future grid of models (\S\ref{future}), and to show applications of this tool to massive young stellar objects (\S\ref{discuss}).

%%%%%%%%%%%%%%%%%%%%%%%%%%%%%%%%%%%%%%%%%%%%%%%%%%%%%%%%%%%%%%%%%%%%%%%%%%%%%%%%%%%%%%%%%%%%%%
\section{Limitations and caveats}
%%%%%%%%%%%%%%%%%%%%%%%%%%%%%%%%%%%%%%%%%%%%%%%%%%%%%%%%%%%%%%%%%%%%%%%%%%%%%%%%%%%%%%%%%%%%%%

\label{limit}

In this section I discuss the main limitations and caveats that apply when using the SED fitting tool. While many of the caveats linked to the radiation transfer models and technique were listed in R06, I discuss here more generally those which directly affect the interpretation of modeling results.

%%%%%%%%%%%%%%%%%%%%%%%%%%%%%%%%%%%%%%%%%%%%%%%%%%%%%%%%%%%%%%%%%%%%%%%%%%%%%%%%%%%%%%%%%%%%%%
\subsection*{Distinguishing formation scenarios - consistency versus proof}
%%%%%%%%%%%%%%%%%%%%%%%%%%%%%%%%%%%%%%%%%%%%%%%%%%%%%%%%%%%%%%%%%%%%%%%%%%%%%%%%%%%%%%%%%%%%%%

\label{proof}

When carrying out modeling of YSO SEDs, \textit{consistency} is sometimes mistaken for \textit{proof}. Simply because a number of SEDs in our (or any other) model SED grid fit an observed SED, this does not prove that any of these models are actually the correct ones for the object in question, only that they are consistent with the observations.\\

This occurs most commonly when attempting to distinguish various scenarios of massive star formation. For example, consider three of the main competing theories of massive star formation, \textit{monolithic collapse} (McKee \& Tan, 2003), \textit{competitive accretion} (Bonnell et al., 2001), and \textit{stellar mergers} (Bonnell et al., 1998). The main qualitative difference between these models are that:

\begin{itemize}
\item in the monolithic collapse model, all the material that will form a massive star is initially gravitationally bound in a dense isolated core
\item in the competitive accretion model, most of the material that will form a massive star is initially unbound, and the accretion onto stellar `seeds' proceeds at a different rate depending on their position in the cluster's gravitational potential.
\item in the stellar mergers model, massive stars form through the merging of intermediate mass stars in embedded clusters.
\end{itemize}

While our models may appear more consistent with the monolithic collapse theory, they are also consistent with the competitive accretion scenario for example, in which the gas and dust will likely be in rotational free-fall collapse at small radii ($<0.1$pc), giving rise to similar geometries to the ones we assume.
Simply because our models are able to successfully fit SEDs of massive young stars does not favour one of the formation scenarios over the other. Indeed, such a conclusion would require a critical comparison of SED models predicted for \textit{all three} scenarios, which cannot be done with this current grid of models. In fact, to date, there are no rigourous predictions for differences in the SED of young stars between the three formation scenarios. \textit{In all three cases}, the massive YSOs will be enshrouded in dust and gas, likely producing `Class 0/I' type SEDs of embedded sources. Note that the presence of disks around massive young stars is not an unambiguous signature of monolithic collapse, as one also expects disk structures to form in the other scenarios (Clark et al., in preparation). In future, it is likely to be the radial density profiles of dust and gas around massive YSOs and the properties (rather than the presence) of the disks that provide clues to the formation scenario. 

%%%%%%%%%%%%%%%%%%%%%%%%%%%%%%%%%%%%%%%%%%%%%%%%%%%%%%%%%%%%%%%%%%%%%%%%%%%%%%%%%%%%%%%%%%%%%%
\subsection*{Evolutionary tracks and indirectly observable parameters}
%%%%%%%%%%%%%%%%%%%%%%%%%%%%%%%%%%%%%%%%%%%%%%%%%%%%%%%%%%%%%%%%%%%%%%%%%%%%%%%%%%%%%%%%%%%%%%

\label{tracks}

When sampling the stellar properties for the models in the R06 grid, we sampled stellar masses and ages randomly, then used evolutionary tracks (Bernasconi \& Maeder, 1996; Siess et al., 2000) to obtain consistent radii and temperatures. However, the stellar mass and age themselves are not parameters in the radiation transfer. When using the SED fitting tool, the stellar parameters that are actually directly derived from the SED are the stellar luminosity and temperature, and the mass and age values assigned to each model are correct \textit{if and only if} the evolutionary tracks are correct.\\

In essence, the SED fitting tool does no more or no less than implicitly determining the stellar luminosity and temperature, and placing these on evolutionary tracks in a Hertzprung-Russel diagram. If the stellar luminosity and temperature are well determined from the SED modeling, then the stellar mass and age will also be well determined assuming those specific evolutionary tracks. If either the stellar luminosity or temperature is not well determined, then this will translate into an uncertainty in the stellar mass and age. Since there is no guarantee that the evolutionary tracks are correct, especially at the high-mass end, (a) the mass and age assigned to a particular radius and temperature may be wrong if the evolutionary tracks are wrong, and (b) there is no guarantee that a given temperature/radius combination is physically realistic if the tracks are wrong.\\

More generally, one can extend this to all parameters which are not directly observable from the SEDs. For example, the current models do not include accretion luminosity in the envelope (see R06 for a discussion of this). $\dot{M}_{\rm env}$ can only be determined because it is directly related to the density profile (which is what can actually be measured from the SED) through the assumption of a free-fall rotational collapse model.

%%%%%%%%%%%%%%%%%%%%%%%%%%%%%%%%%%%%%%%%%%%%%%%%%%%%%%%%%%%%%%%%%%%%%%%%%%%%%%%%%%%%%%%%%%%%%%
\subsection*{Trends and correlations}
%%%%%%%%%%%%%%%%%%%%%%%%%%%%%%%%%%%%%%%%%%%%%%%%%%%%%%%%%%%%%%%%%%%%%%%%%%%%%%%%%%%%%%%%%%%%%%

\label{trends}

\begin{figure}[t]
\begin{center}
\plotone{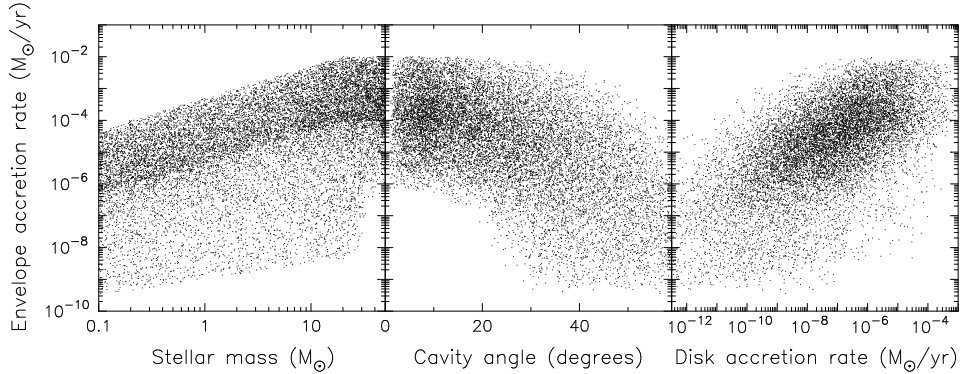}
\caption{Selection of two-dimensional cuts in 14-dimensional parameter space showing artificial correlations between the parameters of the models in the grid. These arise because the parameters were originally not sampled in an unbiased way. \label{parameters}}
\label{default}
\end{center}
\end{figure}

One of the original aims of our work was to search how the evolutionary time-scales of YSOs relate to their stellar mass, and more generally to search for trends between physical parameters, time-scales, spatial locations, and environments. However, we have found that the current grid of model falls short in answering these questions. The sampling of parameter space was deliberately biased in various ways as the computing time required for a completely unbiased grid would have been too large at the time. However, the biases in 14-dimensional parameter space are very difficult to comprehend, and therefore any trend found from SED modeling using the current grid of models may be a classic case of \textit{what you get out is what you put in}. Figure \ref{parameters} shows examples of artificial correlations between parameters, which are directly due to the way that parameter space was sampled in R06. For example, the envelope accretion rate is correlated with the stellar mass and the disk accretion rate, and inversely correlated with the cavity opening angle. We strongly encourage investigators to download the parameter values for all models from the web server\footnote{http://www.astro.wisc.edu/protostars/repository/} in order to check whether any artificial correlations can affect their results, by producing cuts through parameter space as in Figure \ref{parameters}.

%%%%%%%%%%%%%%%%%%%%%%%%%%%%%%%%%%%%%%%%%%%%%%%%%%%%%%%%%%%%%%%%%%%%%%%%%%%%%%%%%%%%%%%%%%%%%%
\subsection*{Multiple sources}
%%%%%%%%%%%%%%%%%%%%%%%%%%%%%%%%%%%%%%%%%%%%%%%%%%%%%%%%%%%%%%%%%%%%%%%%%%%%%%%%%%%%%%%%%%%%%%

\label{multiple}

One major caveat mentioned in R06 was that none of the models in the current grid include binary central sources. However, there is increasing evidence, some of which was presented at this meeting, that multiplicity appears to be unavoidable when looking at massive stars. For example, while the average number of companions in the Orion Nebula Cluster is approximately 0.5 for low-mass stars, it is at least 1.5 for the high mass stars (Preibisch et al., 2001). Therefore, the statement by Mathieu (1994) that ``Binary formation is the primary branch of the star formation process'' is more true than ever, especially at the high-mass end.\\

When designing the current grid of models, we attempted to account for binaries inside a shared envelope by allowing large inner holes in envelopes and disks at all evolutionary stages. However, the central sources in our models are only single stars. While the bolometric luminosity and `average' stellar temperatures can still be recovered from SED modeling, transforming these into a stellar mass and age may be incorrect if one assumes the central source is single when it is in fact multiple.\\

In future, we plan to investigate the effects of binarity and higher multiplicity on the SEDs of young stars. For example, replacing a single source by a close binary system in a YSO will likely only change the apparent stellar luminosity and/or temperature, while wider binaries will likely require a different dust geometry altogether.

%%%%%%%%%%%%%%%%%%%%%%%%%%%%%%%%%%%%%%%%%%%%%%%%%%%%%%%%%%%%%%%%%%%%%%%%%%%%%%%%%%%%%%%%%%%%%%
\subsection*{Confusion}
%%%%%%%%%%%%%%%%%%%%%%%%%%%%%%%%%%%%%%%%%%%%%%%%%%%%%%%%%%%%%%%%%%%%%%%%%%%%%%%%%%%%%%%%%%%%%%

\begin{figure}[t]
\begin{center}
\plotone{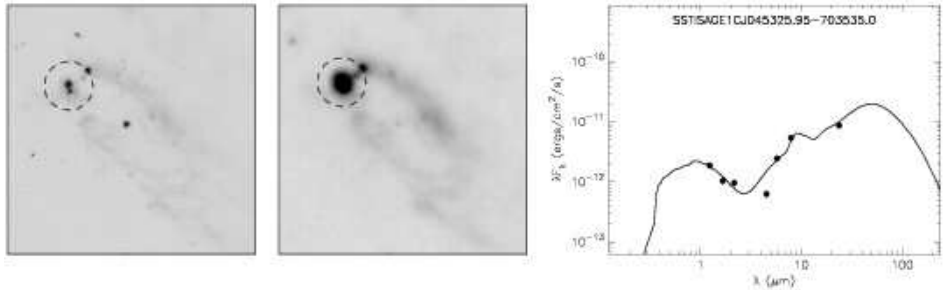}
\caption{\textit{Left and Center} - \textit{Spitzer} IRAC 8.0$\mu$m and MIPS 24$\mu$m observations of a source in the LMC which has mid-infrared colours and spatial location consistent with a YSO (shown circled). \textit{Right} - the SED for this source, which appears unlike conventional `Class I' sources, and appears strongly double peaked. The near-infrared colours are nearly photospheric. While we can find a model from the R06 grid which will provide a good fit to this object (solid line), it is more likely that we are in fact seeing the superposition of a young population dominating the mid- and far-infrared fluxes, and a more evolved population dominating the near-infrared fluxes, especially when considering that the resolution of IRAC at distance of the LMC is of the order of at least 0.5pc.\label{lmcsource}}
\label{default}
\end{center}
\end{figure}

\label{confusion}

A similar problem to multiplicity is that of confusion. Due to the relative rarity of massive YSOs compared to low-mass YSOs, the former are typically seen at kpc distances. At these distances, the resolution of the \textit{Spitzer Space Telescope} at 24$\mu$m for example corresponds to a projected distance of the order of tens of thousands of AU. In dense forming clusters, this means that it is likely for several YSOs to be present in the point spread function (PSF), and even more so at longer wavelengths. At the distance of the LMC, it is almost certain that all point source YSOs contain multiple objects in the PSF (even though the multiple objects may not necessarily be physically associated). An example of an LMC YSO candidate is shown in Figure \ref{lmcsource}, demonstrating that although it is possible to find a model which will correctly reproduce its SED, it is more likely that there are multiple sources within the \textit{Spitzer} PSF.
%%%%%%%%%%%%%%%%%%%%%%%%%%%%%%%%%%%%%%%%%%%%%%%%%%%%%%%%%%%%%%%%%%%%%%%%%%%%%%%%%%%%%%%%%%%%%%
\subsection*{Dust assumptions}
%%%%%%%%%%%%%%%%%%%%%%%%%%%%%%%%%%%%%%%%%%%%%%%%%%%%%%%%%%%%%%%%%%%%%%%%%%%%%%%%%%%%%%%%%%%%%%

\label{dust}

A detailed discussion of existing dust models is beyond the scope of this contribution, but the main problem is the following: although our understanding of dust in the ISM and YSOs is improving, we do not know for certain whether the dust model we assume in the radiation transfer is even close to the true dust opacity. A different dust opacity law would change the dust mass needed to reproduce an observed SED, and therefore the gas mass (assuming a constant dust-to-gas ratio), and more generally will affect the radiation transfer as a whole.

%%%%%%%%%%%%%%%%%%%%%%%%%%%%%%%%%%%%%%%%%%%%%%%%%%%%%%%%%%%%%%%%%%%%%%%%%%%%%%%%%%%%%%%%%%%%%%
\subsection*{Unique fits and data-related issues}
%%%%%%%%%%%%%%%%%%%%%%%%%%%%%%%%%%%%%%%%%%%%%%%%%%%%%%%%%%%%%%%%%%%%%%%%%%%%%%%%%%%%%%%%%%%%%%

\label{unique}

The above limitations relate to the interpretation of SED fitting results from a scientific point of view, but there are also technical issues which can affect results from SED modeling. For example, erroneous fluxes can seriously affect the chances of obtaining a sensible model SED fit to observations. It is therefore very important to ensure that the data is of sufficient quality to allow fitting of model SEDs. This is especially true in this era of large infrared surveys, where the large number of sources makes manual checking of the photometry impossible. However, when extracting a small number of sources from such surveys, the data should be checked before carrying out modeling of the SEDs.\\

It is also important to realise that a unique fit within a given $\chi^2$ tolerance does \textit{not} signify that one has found the only set of parameter values which can reproduce the observations, but merely that one is being too restrictive in what to call a good fit. This can often occur when one specifies very small errors on data-points, or in the presence of a bad data-point.

%%%%%%%%%%%%%%%%%%%%%%%%%%%%%%%%%%%%%%%%%%%%%%%%%%%%%%%%%%%%%%%%%%%%%%%%%%%%%%%%%%%%%%%%%%%%%%
\section{Future work}
%%%%%%%%%%%%%%%%%%%%%%%%%%%%%%%%%%%%%%%%%%%%%%%%%%%%%%%%%%%%%%%%%%%%%%%%%%%%%%%%%%%%%%%%%%%%%%

\label{future}

We are in the process of designing a new grid of model SEDs. The main scientifically important changes over the previous grid of models include (a) a regular and unbiased sampling of parameter space (b) independence of the models from any pre-main sequence evolutionary tracks (c) the inclusion of various dust types, and (d) images at all wavelengths. Points (a), (b), and (c) all address possible limitations that have been discussed in Section \ref{limit}. Once parameter space is regularly sampled in an unbiased way, we will be able to look for trends in the physical conditions in YSOs, for example correlations between envelope dust mass and stellar mass. The models will be defined without a stellar mass and age, since these are not needed for the radiation transfer (only the stellar radius and temperature are needed to define the properties of the central source), and we will leave it to the discretion of users of the models to compute stellar masses and ages using their favourite evolutionary tracks. The inclusion of a few different dust types will allow us to understand how important the choice of a dust model is. In addition to computing this new grid of models, we also plan to carry out a detailed investigation of the effects of multiplicity and confusion on SEDs.

%%%%%%%%%%%%%%%%%%%%%%%%%%%%%%%%%%%%%%%%%%%%%%%%%%%%%%%%%%%%%%%%%%%%%%%%%%%%%%%%%%%%%%%%%%%%%%
\section{Discussion}
%%%%%%%%%%%%%%%%%%%%%%%%%%%%%%%%%%%%%%%%%%%%%%%%%%%%%%%%%%%%%%%%%%%%%%%%%%%%%%%%%%%%%%%%%%%%%%

\label{discuss}

\begin{figure}[t]
\begin{center}
\plotone{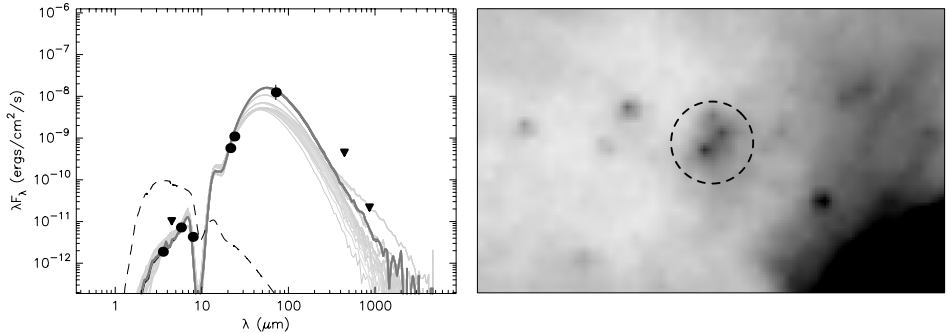}
\caption{\textit{Left} - broadband SED for the embedded protostellar object G34.4+0.23MM (filled circles) shown with several model SEDs providing a good fit (solid lines). \textit{Right} - \textit{Spitzer} IRAC 8.0$\mu$m image of the source (shown circled), which shows two diffuse sources which could be tracing scattered light from a bipolar cavity carved by an outflow.\label{g34.4mm}}
\label{default}
\end{center}
\end{figure}

In this contribution, I have placed a strong emphasis on the limitations of the YSO SED modeling technique we have developed. However, with these limitations in mind, the models and SED fitting tool can be used to determine useful information about massive YSOs.\\

An example of useful quantitative application of the fitting tool to the study of massive YSOs is the determination of bolometric luminosities of massive YSOs by the Red MSX Source (RMS) survey team (Hoare et al., 2004). As discussed in Section \ref{tracks}, parameters such as the stellar mass, age, or the envelope accretion rate are not parameters which relate directly to the SEDs, as opposed to parameters such as the bolometric luminosity and stellar temperature. While luminosities could be determined simply by integrating SEDs and assuming isotropic emission, the strength of the SED fitting technique is that it implicitly takes into account that the emission is not necessarily isotropic (since the models can have non-isotropic emission depending on the dust geometry), and therefore gives a handle on the uncertainties on the bolometric luminosity.\\

Another example, qualitative this time, is that of the embedded protostellar object G34.4+0.23MM (Shepherd et al., 2007). This source is very bright as seen by \textit{Spitzer} at 24$\mu$m, whereas IRAC observations show two faint diffuse sources, shown in Figure \ref{g34.4mm}. By using the SED fitting tool, we found that we can fit the observations well with models of an embedded massive star seen edge-on with bipolar cavities (which tend to scatter light). The two diffuse sources seen at IRAC wavelengths might then correspond to the scattered light from the bipolar cavities. Further support to this interpretation are that a CO outflow is seen emanating from this source, and is aligned with the potential bipolar cavities.\\

In summary, I have provided in this contribution a review of some of the main limitations in our models and SED fitting technique, in particular when applied to massive YSOs, as well as examples of use of the fitting tool to derive quantitative or qualitative information about massive YSOs. The work carried out so far is only the first step towards improving our ability to use SEDs to learn about massive star formation, and we will address a number of limitations with the next grid of models that will be computed at the start of 2008.

\acknowledgements %%% Text of acknowledgements runs on after this command.

I wish to thank Barbara Whitney for comments on the manuscript, and Ian Bonnell for useful discussions. Funding was provided by a Scottish Universities Physics Alliance Studentship.

%%% THE BIBLIOGRAPHY
%%%
%%% CONSULT SECTION 3 OF "INSTRUCTIONS FOR AUTHORS" FOR HOW TO USE NATBIB.
%%% AUTHORS ARE ENCOURAGED TO USE EITHER THE "THEBIBLIOGRAPY" ENVIRONMENT
%%% BY UNCOMMENTING (DELETING THE "%" SYMBOL) THE COMMANDS BELOW, OR BY
%%% USING THE BIBTEX ENVIRONMENT. TO FIND OUT WHICH IS APPLICABLE TO YOUR
%%% CONTRIBUTION, CONSULT THE VOLUME EDITORS FOR YOUR PROCEEDINGS.
%%%

\end{document}